\documentclass[preprint2]{aastex}
\usepackage{psfig}

\journalid{}{}
\articleid{}{}

\slugcomment{Submitted to {\it The Astrophysical Journal (Letters)}}

\shorttitle{RX~J0911+05: A Massive Cluster Lens at z=0.769}
\shortauthors{Kneib, Cohen \& Hjorth}

\begin{document}

\title{
RX~J0911+05:  A Massive Cluster Lens at z=0.769}

\author{Jean-Paul Kneib$^1$}
\affil{Observatoire Midi-Pyr\'en\'ees, 14 Av. E.Belin, 31400 Toulouse,
France}
\email{kneib@ast.obs-mip.fr}

\author{Judith G. Cohen}
\affil{Palomar Observatory, Mail Stop 105-24,
        California Institute of Technology, Pasadena, CA \, 91125, USA}
\email{jlc@astro.caltech.edu}

\author{Jens Hjorth}
\affil{Astronomical Observatory, University of Copenhagen,
Juliane Maries Vej 30, DK-2100 Copenhagen, Denmark}
\email{jens@astro.ku.dk}

\altaffiltext{1} {Visiting astronomer,
of W.M. Keck Observatory, which is operated as a scientific
partnership among the California Institute of Technology, the
University of California and the National Aeronautics and Space
Administration.  The Observatory was made possible by the generous
financial support of the W.M. Keck Foundation.}

\begin{abstract}
  We report the detection of a massive
  high-redshift cluster of galaxies near the quadruple quasar
  RX~J0911+05,  using the LRIS instrument on the Keck-II telescope.
  The cluster is found to have a mean redshift of
  $\bar{z} =0.7689\pm 0.002$ and a velocity dispersion of $\sigma =
  836^{+180}_{-200}$ km~s$^{-1}$,
  based on redshift measurements for 24 member galaxies.
  This massive high-redshift cluster is the origin of the
  unusually large external shear required by lensing models of
  the quadruple quasar system.  We predict the
  expected time delay depending on the exact contribution of the
  cluster.  A measurement of the time delay and further deep lensing 
  and X-ray observations will unravel
  useful properties of this serendipitously discovered high-redshift
  cluster, and may put interesting cosmological constraints on $H_0$.

Subject headings: galaxies: clusters: general ---  gravitational lensing ---
galaxies: clusters: individual (RX~J0911.4+0551)
\end{abstract}

\section{Introduction}
Multiply imaged quasars are among the most promising objects for
measuring distances and hence cosmological parameters (Refsdal 1964;
Blandford \& Narayan 1992). That promise is finally starting to be
realized after time delays have been measured in several systems in
recent years (e.g., Schechter et al.\ 1997; Kundi\'c et al.\ 1997c;
Hjorth et al.\ 1999). Contrary to local estimates of the Hubble
parameter, multiply imaged quasars with time delays can be used to measure
the Hubble parameter on a {\it cosmological}\/ scale and hence are
insensitive to local bulk flows.

Until recently,
simple mass distributions have been used to model multiply imaged
QSOs.  However, the rapid development of observing techniques, and the
availability of powerful observatories like the {\sl Hubble Space
Telescope\/} and the 8--10 m class telescopes have shown that most of
the image configurations are more complex than allowed by simple
models.  As an example, the famous double QSO~0957+561, first thought
to be produced by a single galaxy, is now known to be a complex system
involving a massive lens plus one or more galaxy clusters (Bernstein
\& Fisher 1999 and references therein).  In parallel to the 
surge in observational efforts, the modeling of galaxies and their
surroundings has greatly improved and is starting to provide very
important insights into the distribution of dark matter in galaxies
(e.g.\ Hjorth \& Kneib 2000; Maller et al.~2000; Kochanek et al.~2000).

A major uncertainty in the mass models comes from the surroundings of
the lens, in the form of ``external shear'', e.g., due to either
nearby galaxies, a galaxy group or a cluster close to the line of
sight to the QSO (Keeton, Kochanek \& Seljak 1997, Kundic et al.~1997b).
It is important to underline that a reliable mass model requires not
only $\gamma$, the gravitational external shear, but also the external
surface mass density $\kappa$, producing the measured external shear.
Therefore, a measurement of the mass distribution in the vicinity of
the lensed quasar is {\it mandatory} in putting stringent constraints
on cosmological parameters.

The quadruple lens system RX~J0911+05, is potentially one of the most
valuable systems for cosmological studies.  It was identified by Bade
et al.\ (1997) as a multiply imaged radio quiet QSO at $z=2.80$,
selected as an AGN candidate from the ROSAT All-Sky Survey.  In
high-resolution NOT and NTT images (Burud et al.\ 1998), RX~J0911+05
was soon found to consist of four QSO components and an elongated
lensing galaxy, confirmed by subsequent {\sl HST\/} imaging. Regular
monitoring of RX~J0911+05 with the Nordic Optical Telescope is 
expected to give a time-delay to about 10 percent (Hjorth et al.~2000) 
Moreover, recently, the redshift
of the main lens was measured at $z=0.769$ with the Keck telescope
(Falco, Davis \& Stern, 1999, private communication).
The unusual image configuration of RX~J0911+05 requires a large
external shear ($\gamma_{\rm min}=0.15$) to be included in the lensing
potential in order to reproduce the complex geometry observed in
RX~J0911+05. The probable source of this shear is a galaxy cluster
located about 38\arcsec\, from RX~J0911+05 as suggested by Burud 
et al.~(1998). Such a cluster may contribute to the X-ray emission observed
by ROSAT.  The color of the presumed cluster galaxies indicates a
redshift in the range of 0.6--0.8, consistent with the determined
spectroscopic redshift of the main lens galaxy.  

In this Letter we present the results of spectroscopic observations
of RX J0911+05 conducted at the Keck Observatory  using the LRIS
instrument.  Section 2 summarizes the photometric and spectroscopic
observations.  Section 3 presents the constraints on the mass model of
the high-redshift cluster of galaxies detected near the quadruple
lens.  A discussion of the results is presented in section 4.
Throughout this paper we use a Hubble parameter $H_0 = 50 h_{50}^{-1}$
km~$s^{-1}$~Mpc$^{-1}$, $\Omega_{\rm M} = 0.3$ and ${\Omega_\Lambda} =
0.7$, which gives a physical scale of $10.37 h_{50}^{-1}$ kpc arcsec$^{-1}$
at the cluster redshift ($z=0.769$).

\section{Observations}

We observed RX~J0911+05 during the December 7--8, 1999 and March 6--7,
2000 nights with the LRIS instrument (Oke et al.~1995) mounted on Keck
II. Imaging was done in B, R and I and calibrated using photometric
standards PG2213-006 and other stars in the SA92 field (Landolt 1992).
Spectroscopic observation were conducted with the 400/8500 grating
blazed at 7200~\AA\ which gives a dispersion of 1.84~\AA/pixel.
Standard {\sl Figaro} and {\sl IRAF} pipeline reduction packages where
used to reduce the photometric and spectroscopic data. The Journal of
these observations is given in Table \ref{tab:journal}. A color image
made of the B, R and I images is presented in Figure \ref{fig:color}.

We exposed 2 MOS masks, the first mask had 24 slitlets, and the second 30
slitlets. Object selection was limited in magnitude to $I_{AB} < 22.5$
and generally followed the color selection of Burud et al.~(1998).  In
the first mask, a total of 25 objects were observed (there were two objects
in one slitlet); 3 stars and 2 bright galaxies were used to allow a
correct alignment of the mask on the objects. Similarly for the second
mask, we targeted 30 galaxies, and used 3 stars for the mask
alignment.

\begin{figure}
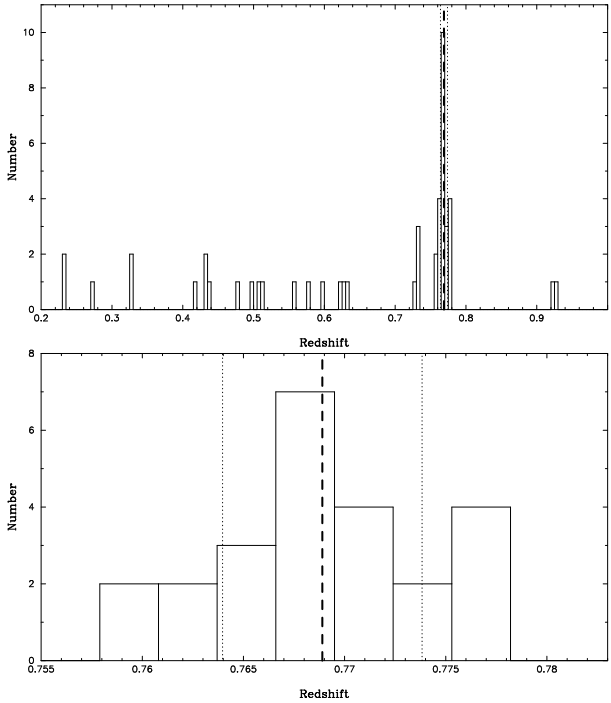

\begin{minipage}{8.cm}
\psfig{file=fig2a.ps,width=\textwidth}
\end{minipage}
\begin{minipage}{8.cm}
\psfig{file=fig2b.ps,width=\textwidth}
\end{minipage}
\caption{{\bf  (left)} Redshift histogram of the redshift survey.
{\bf (right)} Redshift histogram of the galaxy distribution at the cluster
redshift.  The dashed lined corresponds to the mean redshift of the
cluster distribution, the dotted line to the standard dispersion of
836$^{+180}_{-200}$~km~s$^{-1}$.  }
\label{fig:zhist}
\end{figure}

Redshifts were estimated independently by JPK and JGC, by simple line
fitting techniques, as well as using the cross-correlation package
$RVSAO\ V2.0$ in $IRAF$ (Kurtz \& Mink 1998).  Redshift
identifications are displayed on Figure \ref{fig:color}.  Table
\ref{tab:redshift} summarizes the photometry and the redshift of the
cluster members.  Figure
\ref{fig:zhist} shows the redshift histogram of the entire sample and the
cluster sample.  The majority of the cluster member spectra are
typical of an old stellar population. Only 3 out of 23 spectra show a
strong [\ion{O}{2}] emission line (one of them being an AGN), and 4
others show a weak [\ion{O}{2}] line.  Although our target selection
was clearly biased to have a large number of early type galaxies, it
clearly shows that these galaxies belong to an evolved cluster. Note
that the small group of apparently redder objects detected some
10\arcsec\, SW from the lens (Burud et al.\ 1998) is found at the same
redshift as the cluster.

\section{Cluster Dynamics and Mass Model}

Including the redshift of the lensing galaxy (Falco, Davis \& Stern~
1999, private communication) we have redshift measurements of 24
cluster members with a mean redshift $\bar{z}=0.7689 \pm 0.002$. The
velocity dispersion of the 24 galaxies is $\sigma_{\rm los} = 836
^{+180}_{-200}$~km~s$^{-1}$. If one keeps only the 14 high quality
spectra, we estimate the mean redshift of $\bar{z}=0.7692 \pm 0.004$
and a velocity dispersion of $\sigma_{\rm los} = 832
^{+175}_{-250}$~km~s$^{-1}$. The two estimates are in good agreement,
so we will retain the first one in the following discussion.  We
used the biweight estimator to measure the cluster redshift and its
dispersion. The errors were estimated by using the bootstrap algorithm
for the mean redshift and the jacknife algorithm for the velocity
dispersion (Beers, Flynn \& Gebhardt 1990).

We computed the harmonic radius $R_h$ (e.g., Nolthenius \& White 1987)
from our spectroscopic cluster members as
\begin{equation}
\displaystyle
R_h = D_A(\bar{z}) {\pi\over 2}{N_m(N_m-1)\over 2}
\left(\Sigma_i\Sigma_{j>i}\theta_{ij}^{-1}\right)^{-1},
\end{equation}
where $\theta_{ij}$ is the angular distance between galaxies $i$ and
$j$, $N_m$ is the number of cluster members, and $ D_A(\bar{z})$ is
the angular diameter distance at the mean cluster redshift $\bar{z}$.
The cluster virial mass can then be estimated as
\begin{equation}
M_V = { 6 \sigma^2 R_h \over G}.
\end{equation}
We found an harmonic radius of $R_h = 632 h_{50}^{-1}$ kpc and a mass
$M_V = 6.2^{+2.9}_{-2.6} \,10^{14}$ M$_\odot$.  Assuming that the
cluster follows the $\sigma$--$T_X$ relation ({\it e.g.} 
Girardi et al.~1996), we derive an X-ray temperature of 
$T_X= 4.5 \pm 1.2$ keV.

We have thus identified a new massive cluster at high redshift. This
cluster is not as massive as the cluster MS1054$-$03 (Tran et
al.~1998; Hoekstra et al.~2000) but is similar to MS1137.5+6625
($\sigma = 884$~km~$s^{-1}$ as measured by Donahue et al.~2000).

Assuming a singular isothermal sphere centered on the brightest galaxy
of the cluster, the computed velocity dispersion translates into
a gravitational shear of $\gamma =
\kappa =0.11^{+0.04}_{-0.03}$ for $(\Omega, \lambda)=(1, 0)$ and
$\gamma = \kappa =0.13^{+0.04}_{-0.03}$ for $(\Omega, \lambda)=(0.3,
0.7)$ at the location of the multiple quasar.
This quick estimate of the external shear is close to the
value found in the modeling of the lens plus an external shear
(Burud et al.~1998).

\begin{figure}
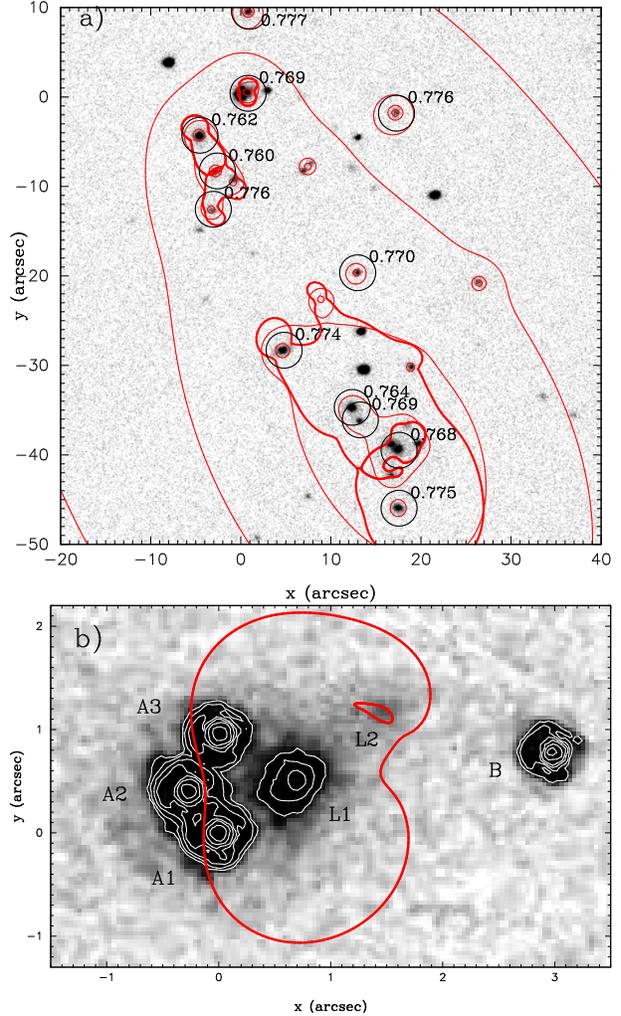

\begin{minipage}{8.cm}
\psfig{file=fig3a.ps,width=\textwidth}
\end{minipage}
\begin{minipage}{8.cm}
\psfig{file=fig3b.ps,width=\textwidth}
\end{minipage}

\caption{{\bf  a)} Mass distribution of the cluster
 (thin lines) overlaid on the K-band image (Courbin, private
 communication). The thick lines are the critical lines at the
 redshift $z=2.80$ of the quasar.  {\bf b)} HST/NICMOS-2 F160W
 archival image of the quadruple quasar RX~J0911+05 (PI: Falco). The
 main lens is composed of two galaxies $L1$ and $L2$. The critical lines
 (thick lines) correspond to the redshift $z=2.80$ of the quasar.
 Image A1 is the reference coordinates (0, 0). North is up, East to
 the left. }
\label{fig:model}
\end{figure}

Having identified the cluster redshift and derived an estimate of the 
galaxy velocity dispersion, we can go a step further and try to model 
the system modeling mass distribution of the cluster and its galaxies in 
a similar way as for Abell 2218 (Kneib et al.~1996).  
The cluster galaxies are modeled as truncated isothermal
spheres, with a velocity dispersion and truncation radius scaled with
the galaxy luminosity. The center position of the cluster is set to
the position of the brightest galaxy and the velocity dispersion as
given by our spectroscopic survey. The constraints are the quasar
positions and flux ratios (although we only give a relatively small
weight to the flux ratio constraints due to possible microlensing
(Burud et al.~1998),
the position, orientation, and ellipticity of the main lens system as
measured on the HST/NICMOS-2 image (PI: Falco). The fiducial best-fitting 
model found is displayed in Figure \ref{fig:model}, where we show the 
surface mass density of the model.
Note that the lens model favors an elongated mass distribution of the
cluster aligned with the cluster light distribution. The time delay
expected between images $A$ and $B$ is strongly dependent on the exact
morphology and mass distribution of the cluster; we found that
$190 < \Delta t\ {\rm (days)}\ < 260 h_{50}^{-1}$. The lower time-delay 
value corresponds to a
more massive cluster.  Clearly it would be very important to further
constrain the mass distribution of the cluster before trying to get a
constraint on the Hubble parameter.

\section{Discussion}

We have identified a massive high-redshift cluster of galaxies
at $\bar{z}=0.769\pm 0.002$ responsible for the very large external shear ($\gamma
\sim 0.15$), affecting the lens potential of the quadrupole quasar
RX~J0911+05. The measured velocity dispersion based on 24 members
leads to a velocity dispersion of $\sigma=836 ^{+180}_{-200}$~km~$s^{-1}$.
Using these results we present a new mass model for this lens
which includes the contribution of a cluster and the cluster galaxies
in a similar way as Kneib et al.~(1996). The predicted time delay is
estimated to be $190 < \Delta t\ {\rm (days)}\ < 260 h_{50}^{-1}$, 
depending on the exact mass distribution of the cluster.

Further observations using the WFPC2 on board HST will allow
us to directly measure the predicted cluster weak shear from the
distortions of faint background galaxies; X-ray observations conducted
with the new X-ray satellites ({\sl Chandra} or {\sl XMM-Newton}) will
permit to study the gas properties of this distant
cluster in detail. Such complementary data will be essential in producing a
reliable mass model of the system which, in combination with the
foreseen measurement of the time delay, may provide the best
constrained mass model for a multiple QSO, hence leading to an
accurate cosmological estimate of the Hubble parameter.

This serendipitous cluster discovery is very interesting because it
demonstrates that multiple quasars with large separation are efficient
in revealing high-redshift collapsed structure. Indeed, there is a
growing number of such multiple quasars where a cluster or group as
been detected (e.g., Q0957: Bernstein \& Fischer 1999; the Cloverleaf:
Kneib et al.~1998a, 1998b, MG2016+116 Hattori et al.~1997; Benitez et
al.~1999; Soucail et al.~2000, PG1115+080: Schechter et al.~1997,
Kundic et al.~1997a). Therefore, multiple QSO system, constitute a
useful way to discover and study high-redshift groups and clusters
along their line of sight.

\acknowledgements
 
 Part of the work presented is based on observations made with the
 NASA/ESA Hubble Space Telescope, obtained from the data archive at
 the Space Telescope Science Institute. STScI is operated by the
 Association of Universities for Research in Astronomy, Inc. under
 NASA contract NAS 5-26555. The entire Keck/LRIS user community owes a
 huge debt to Jerry Nelson, Gerry Smith, Bev Oke, and many other
 people who have worked to make the Keck Telescope and LRIS a reality.
 We are grateful to the W. M. Keck Foundation, and particularly its
 late president, Howard Keck, for the vision to fund the construction
 of the W. M. Keck Observatory.\\
 It is a pleasure to acknowledge the
 efficient and friendly support of the Keck Observatory staff.  We
 also acknowledge useful discussions with C. Kochanek, F. Courbin, and
 I. Burud.  This research was supported by CNRS/INSU for JPK and by
 the Danish Natural Science Research Council (SNF) for JH.

\begin{deluxetable}{cccc}
\tablecolumns{4}  
\tablewidth{0pc}
\tablecaption{Journal of observations of RX~J0911+05 
done  at Keck II using LRIS on December 7-8, 1999 and on March 6-7, 2000.}
\label{tab:journal}
\tablehead{
\colhead{Filter or} & \colhead{Exposure} & \colhead{Pixel} &
 \colhead{Seeing} \\
\colhead{$\lambda$-range} & \colhead{Time (sec)} &
 \colhead{($\arcsec$) - (\AA)} & \colhead{FWHM ($\arcsec$)}
} 
\startdata
 I-band & 300  & 0.215  & 1.1 \\
 R-band & 140  & 0.215 & 1.1 \\
 B-band & 200  & 0.215   & 1.2 \\
\noalign{\smallskip}
\tableline
\noalign{\smallskip}
 5000--9000~\AA  &3000  & 0.215 - 1.84 & 1.1 \\
\noalign{\smallskip}
\tableline
\noalign{\smallskip}
 5000--9000~\AA  &3000  & 0.215 - 1.84 & 0.9 \\
\enddata  
\end{deluxetable}

\begin{deluxetable}{llllllccc}
  \tablecolumns{9} \tablewidth{0pc} \tablecaption{ Catalogue of the
  cluster galaxies in the field of RX~J0911+05.  
  }
  \tablecomments{$R$ is the
  cross-correlation parameter. Type and quality classification is
  similar to the one used by Cohen et al.~(1999). L1 is the main galaxy
  lens.  }
\label{tab:redshift}
\tablehead{
\colhead{ID} & \colhead{$\alpha$} & \colhead{$\delta$} &  
\colhead{I$_{AB}$} & \colhead{$z$} & \colhead{$\Delta z$} & 
\colhead{Line} & \colhead{Cross-correlation} & \colhead{Object} \\
\colhead{} & \colhead{$ (J2000) $} & \colhead{$(J2000) $} & \colhead{}  
&\colhead{}  &\colhead{}  & \colhead{Identification} & 
\colhead{template -$R$} & \colhead{type-quality} 
}
\startdata
G1 & 09:11:26.46 & +05:50:14.5  & 19.99 & 0.7682 & 0.0002 &
H/K/br/G & k0V/g6V - 3.03 & \cal{A}-1 \\
G2 & 09:11:27.29 & +05:50:25.4  & 20.82 & 0.7738 & 0.0003 &
[\ion{O}{2}]/Balmer/H/K/G & spec-em - 3.92 & {\cal C}-1 \\
G3 & 09:11:27.92 & +05:50:49.5  & 20.90 & 0.7618 & 0.0003 &
H/K/br/G& g6V - 3.43 & {\cal A}-1 \\
G4 & 09:11:21.17 & +05:49:39.8  & 20.94 & 0.7693 & 0.0004 &
H/K/br/G & gell - 5.40 & {\cal A}-1 \\
G5 & 09:11:21.40 & +05:48:03.8  & 21.24 & 0.7697 & 0.0003 &
H/K/br/G & gell - 7.50 & {\cal A}-1 \\
G6 & 09:11:28.36 & +05:51:07.9  & 21.30 & 0.7650 & 0.001 & 
wk[\ion{O}{2}]/K/H/G & --- & {\cal C}-1 \\
G7 & 09:11:26.43 & +05:50:07.7  & 21.65 & 0.7754 & 0.0005 &
H/K/br/G & gell - 4.49 & {\cal A}-1 \\
G8 & 09:11:26.45 & +05:50:52.0  & 21.71 & 0.776 & 0.001 & 
wk[\ion{O}{2}]/H/K/br & --- & {\cal C}-1 \\
G9 & 09:11:26.39 & +05:49:52.7  & 21.72 & 0.7674 & 0.0003 &
H/K/br/G & f6V - 3.73 & {\cal A}-1 \\
G10 & 09:11:22.93 & +05:52:40.3  & 21.87 & 0.7628 & 0.0004 & 
MgII/[\ion{O}{2}]/H/K & --- & {\cal C}-1 \\
G11 & 09:11:22.76 & +05:48:29.9  & 21.89 & 0.7692 & 0.0002 &
H/K/br/G & gell - 5.90 & {\cal A}-1 \\
G12 & 09:11:30.61 & +05:50:17.1  & 21.94 & 0.766 & 0.001 & 
H/K/br/G & --- &         {\cal A}-2 \\
G13 & 09:11:31.59 & +05:50:37.6  & 21.98 & 0.769 & 0.001 & 
H/K/G & --- &            {\cal A}-3 \\
G14 & 09:11:27.79 & +05:50:45.5  & 22.04 & 0.760 & 0.001 & 
H/K/br & --- &           {\cal A}-3 \\
G15 & 09:11:31.22 & +05:51:26.6  & 22.05 & 0.7745 & 0.0005 & 
wk[\ion{O}{2}]/H$\delta$ & --- & {\cal E}-3 \\
G16 & 09:11:27.82 & +05:50:41.2  & 22.05 & 0.7758 & 0.001 & 
wk[\ion{O}{2}]/K/H/G & --- & {\cal C}-3 \\
G17 & 09:11:27.55 & +05:51:03.4  & 22.10 & 0.7767 & 0.0003 &
H/K/br/G & g6III - 4.01 & {\cal A}-1 \\
G18 & 09:11:24.17 & +05:49:35.2  & 22.16 & 0.760 & 0.001 & 
[\ion{O}{2}]/H/K/br & --- & {\cal E}-3 \\
G19 & 09:11:26.78 & +05:50:19.0  & 22.25 & 0.7640 & 0.0002 &
H/K/br/G & g6III - 6.17 & {\cal A}-1 \\
G20 & 09:11:26.72 & +05:50:17.6  & 22.25 & 0.7694 & 0.0006 &
H/K/br/G & gell - 3.65 & {\cal A}-1 \\
G21 & 09:11:26.74 & +05:50:34.1  & 22.30 & 0.7705 & 0.0005 & 
H/K/br/G & --- &         {\cal A}-3 \\
G22 & 09:11:27.98 & +05:49:18.9  & 22.35 & 0.770 & 0.002 & 
H/K/br/G & --- &         {\cal A}-3 \\
G23 & 09:11:30.57 & +05:51:13.4  & 22.42 & 0.770 & 0.002 & 
H/K/br/G & --- &         {\cal A}-3 \\
L1  & 09:11:27.56 & +05:50:54.21 & 22.4  & 0.769 & \nodata &
\nodata & \nodata & \nodata  \\

\enddata
\end{deluxetable}


\begin{figure}
\caption{Keck/LRIS BRI image ($4.75\arcmin\times6.15\arcmin$ field)
 showing the central region of the RX~J0911+05 cluster.  Circled
 objects were observed in MOS mode, the redshift identification is
 noted on the side of the circle.  The zoomed image corresponds to the
 HST-NICMOS observation of RX~J0911+05 multiple quasar.  }
\label{fig:color}
\end{figure}

\end{document}